%% file: main.tex
\documentclass[letterpaper,twocolumn,10pt]{article}

\usepackage{usenix-2020-09}
\usepackage{prp-macros}

\usepackage[ruled,linesnumbered]{algorithm2e}

\SetCommentSty{algocomment}

\usepackage{colortbl}
\usepackage{listings}
\usepackage{makecell}
\usepackage{multirow}
\usepackage{subcaption}
\usepackage{textcomp}
\usepackage{tikz}
\usepackage{times}
\usepackage{inconsolata}

\definecolor{scheme1}{rgb}{1.0,0.4,0.4}
\definecolor{scheme2}{rgb}{0.29,0.63,0.45}
\definecolor{scheme3}{rgb}{0.2,0.6,1.0}
\definecolor{scheme4}{rgb}{0.3,0.3,0.3}

\usepackage[small,compact]{titlesec}

\makeatletter
\renewcommand{\sectionautorefname}{\S\@gobble}
\renewcommand{\subsectionautorefname}{\S\@gobble}

\makeatother

\lstdefinestyle{cpp}{
    keywordstyle=\bfseries\color{scheme2},
    commentstyle=\itshape\color{scheme1},
    identifierstyle=\color{scheme4},
    stringstyle=\color{scheme3}
}

\lstdefinestyle{py}{
    keywordstyle=\bfseries\color{scheme1},
    commentstyle=\itshape\color{scheme2},
    identifierstyle=\color{scheme3},
    stringstyle=\color{scheme4}
}

\definecolor{boxhead}{rgb}{0.23,0.47,0.84}
\definecolor{boxsubtitle}{rgb}{0.4,0.4,0.4}
\tikzstyle{background rectangle}=[thin,draw=black]

\setcellgapes[t]{0.5pt}
\setcellgapes[b]{0pt}

\usepackage{caption}
\usepackage[compact]{titlesec}

\newcommand{\sys}{\textsc{Faabric}\xspace}
\newcommand{\func}{Granule\xspace}
\newcommand{\funcs}{Granules\xspace}
\newcommand{\group}{Granule group\xspace}
\newcommand{\groups}{Granule groups\xspace}
\newcommand{\cp}{control point\xspace}
\newcommand{\cps}{control points\xspace}
\newcommand{\diff}{byte-wise diff\xspace}
\newcommand{\diffs}{byte-wise diffs\xspace}

\newcommand{\code}[1]{{\small\texttt{#1}}\xspace}

\newcommand{\todo}[1]{\textbf{\color{scheme1}[\textsc{todo:} #1]}}
\newcommand{\gap}{\textcolor{red}{\textbf{[...]}}\xspace}

\begin{document}

\hypersetup{
    colorlinks=true,
    linkcolor=scheme2,
    filecolor=scheme1,
    citecolor=scheme1,
    urlcolor=scheme2
  }

\title{\sys: Fine-Grained Distribution of Scientific Workloads in the Cloud}

\author{
    {\rm Simon Shillaker}\\
    Imperial College London \and
    {\rm Carlos Segarra}
    \\ Imperial College London \and
    {\rm Eleftheria Mappoura}
    \\ Imperial College London \and
    {\rm Mayeul Fournial}
    \\ Imperial College London \and
    {\rm Llu\'is Vilanova}
    \\ Imperial College London \and
    {\rm Peter Pietzuch}
    \\ Imperial College London
}

\maketitle

\input{abstract}

\input{sections/intro}
\input{sections/background}
\input{sections/section3}
\input{sections/section4}
\input{sections/section5}
\input{sections/evaluation}

\input{sections/related}
\input{sections/conclusion}

{\footnotesize\bibliographystyle{plain}
    \bibliography{main}}

\end{document}

%% file: abstract.tex

\begin{abstract}

With their high parallelism and resource needs, many scientific applications benefit from cloud deployments. Today, scientific applications are executed on dedicated pools of VMs, resulting in resource fragmentation: users pay for underutilised resources, and providers cannot reallocate unused resources between applications. While serverless cloud computing could address these issues, its programming model is incompatible with the use of shared memory and message passing in scientific applications: serverless functions do not share memory directly on the same VM or support message passing semantics when scheduling functions dynamically.

We describe \sys, a new serverless cloud runtime that transparently distributes applications with shared memory and message passing across VMs. \sys achieves this by scheduling computation in a fine-grained (thread/process) fashion through a new execution abstraction called \emph{\funcs}. To support shared memory, \funcs are isolated using WebAssembly but share memory directly; to support message passing, \funcs offer asynchronous point-to-point communication. \sys schedules \funcs to meet an application's parallelism needs. It also synchronises changes to \func's shared memory, and migrates \funcs to improve locality.


\end{abstract}




%% file: sections/intro.tex

\section{Introduction}
\label{sec:intro}

Cloud computing offers on-demand access to plentiful resources, making it an attractive choice for highly-parallel scientific applications in hydrodynamics~\cite{OpenFoamGithub}, genomics~\cite{GenomicsCloud}, and epidemiology~\cite{CovidSimGithub}. Such applications often employ parallel programming models that use multi-threading with \emph{shared memory} (\eg OpenMP~\cite{OpenMPWebsite}) and distributed processing with \emph{message passing} (\eg MPI~\cite{MPIWebsite}).

Cloud providers have introduced platforms, \eg AWS Batch~\cite{AWSBatchWebsite} and Azure Batch~\cite{AzureBatchWebsite}, that support such workloads. These platforms typically dedicate a pool of virtual machines~(VMs) to execute a queue of jobs~\cite{AzureBatchWebsite, AWSBatchWebsite}. This allows such platforms to support shared memory and message passing applications: the provider schedules jobs on one or more VMs according to the requested parallelism (\eg MPI's world size or OpenMP's \code{OMP\_NUM\_THREADS}). Shared memory applications are normally executed on a single VM with many CPU cores and large memory sizes.

Deploying scientific applications on a fixed set of VMs, however, leads to resource fragmentation: if a job does not use all available resources or cannot be bin-packed onto available VMs, users pay for idle or under-utilised VMs; providers also cannot exploit a user's idle resources, diminishing the efficiency of their infrastructure~\cite{AWSBatchWebsite, AzureBatchWebsite}.

To achieve higher utilisation, serverless computing~\cite{AWSLambdaWebsite, AzureFunctionsWebsite, GoogleCloudFunctionsWebsite} distributes workloads as fine-grained \emph{functions}, which execute on large VM clusters, isolated through lightweight mechanisms (\eg containers~\cite{OpenFaaSWebsite}, micro-VMs~\cite{Firecracker:NSDI:2020, AWSLambdaWebsite}, or WebAssembly~\cite{WebAssembly:PLDI:2017, Faasm:ATC:2020}). Serverless computing thus reduces under-utilisation by distributing workloads at a fine granularity across machines~\cite{AWSLambdaWebsite, AzureFunctionsWebsite, GoogleCloudFunctionsWebsite}. By making more fine-grained decisions about how to allocate each VM's resources, a provider achieves higher utilisation and thus lower per-tenant costs.


The programming models used in scientific applications, namely shared memory and message passing~\cite{DualityOfOS:1979}, make it challenging to support such applications in serverless environments: (1)~\emph{shared memory} applications rely on threads for parallelism, requiring state to be shared: threads must access the same address space, which is not possible among serverless functions running in different containers, potentially on different VMs. In addition, when deployed in a serverless setting, multi-threaded code is restricted to the parallelism available within a function's isolation boundary, \eg a single container; and (2)~\emph{message passing} applications need a fixed-size pool of stateful processes to support consistent synchronous communication -- this is unavailable in a serverless setting, in which functions tend to be short-lived, stateless and only communicate through storage~\cite{OneStepForwardTwoStepsBack:2019}.







We describe \textbf{\sys}, a new serverless cloud runtime that executes scientific applications with shared memory and message passing. For this, \sys exploits the new abstraction of \emph{\funcs}, which allow for thread- and process-granular scheduling. \sys does not require changes to existing parallel programming models: it transparently executes applications that use OpenMP API~\cite{OpenMPWebsite} for shared memory or MPI API~\cite{MPIWebsite} for message passing. It achieves this through the following contributions:

\mypar{(1)~Supporting multi-threading/processing via \funcs} \sys executes applications as batches of distributed \emph{\funcs}~(\autoref{sec:funcs:funcs}) running on shared VMs. A \func can share memory with other \funcs to offer thread semantics, or have private memory for process semantics. \funcs can be spawned (or migrated) based on a \emph{snapshot} taken from a parent \func. \sys uses WebAssembly~\cite{Wasm:2017} to isolate \funcs and take snapshots of a \func{}'s state: its shared/private memory, message queues, address information and execution state, \eg stack pointers and function tables.

\mypar{(2)~Transparent distribution of \funcs} \sys allows the provider to distribute computation using \funcs. A scheduler in \sys can choose to spawn or migrate \funcs across VM. It makes scheduling decisions when \funcs reach \emph{\cps}, which are triggered by system calls and calls to parallel APIs. At each \cp, \sys may spawn new \funcs to add a logical thread or process to the application, increasing parallelism; or migrate existing \funcs, \eg to increase locality of execution.

\mypar{(3)~Distributed synchronisation of address spaces} To support shared memory programming, \sys must offer sequential consistency within each \func, provide distributed synchronisation primitives, \eg mutexes and barriers, and synchronise distributed writes to shared address spaces~(\autoref{sec:memory}). \funcs synchronise writes to the address space by building lists of \emph{\diffs}. Each \func maintains a record of writes to shared memory pages, performs byte-wise comparisons against its parent snapshot and propagates changes back to a main \func via \diffs. To support updates to shared variables across \funcs, \diff specify a \emph{merge} operation, \eg summation over shared variables.

\mypar{(4)~Asynchronous messaging through \groups} To support message passing between \funcs, \sys organises them into \emph{\groups}. Each \func is assigned an index within the group and can send/receive messages to/from \funcs in the group. \sys maintains a set of queues for each \func to buffer incoming messages, thus sending/receiving messages asynchronously without the need for \funcs to have been scheduled. This prevents message loss during \func migration. \sys implements common collective communication operations, such as all-reduce~\cite{MPIAllReduce}. Its implementation uses fast in-memory message exchange between co-located \funcs on the same VM.

\tinyskip

\noindent
In our evaluation, we use \sys to execute scientific applications implemented using OpenMP and MPI (LAMMPS~\cite{LAMMPSWebsite}, and the ParRes kernels~\cite{ParResKernelsGithub}) and compare them to native OpenMP and OpenMPI. When executing a queue of 100 applications on a 32~VM cluster, \sys can reduce makespan by up to 23\% thanks to its granular scheduling of threads and processes.



%% file: sections/background.tex

\section{Scientific Applications in Cloud}
\label{sec:background}

\begin{table}[t]
  \centering
  \footnotesize
  \begin{tabular}{p{2.5cm}llcc}
    \toprule
    {\textbf{Domain}} & \textbf{Name} & \textbf{Language} & \textbf{SM} & \textbf{MP}\\
    \midrule
    \multirow{2}{*}{\makecell{Molecular dynamics}} & LAMMPS~\cite{LAMMPSGithub} & C++ & \xmark & \cmark \\
                                     & MDAnalysis~\cite{MDAnalysisGithub} & Python & \cmark & \xmark \\
    \midrule
    \multirow{2}{*}{\makecell{Bio-informatics}} & BioPython~\cite{BioPythonGithub} & Python & \cmark & \xmark \\
                                     & gatk~\cite{gatkGithub} & Java & \cmark & \xmark \\
    \midrule
      \multirow{2}{*}{\makecell{Fluid dynamics}} & OpenFOAM~\cite{OpenFoamGithub} & C++ & \xmark & \cmark \\
                                     & SU2~\cite{SU2Github} & C++ & \cmark & \xmark \\
    \midrule
      \multirow{2}{*}{\makecell{Deep learning}} & OpenCV~\cite{OpenCVGithub} & C++ & \xmark & \cmark \\
                                     & Tensorflow~\cite{TensorFlowGithub} & Python & \cmark & \xmark \\
    \bottomrule
  \end{tabular}
  \caption{Github's most-starred projects in scientific domains use shared
  memory~(SM) or message passing~(MP)}
  \label{tab:libraries}
\end{table}

Next, we outline the benefits and challenges associated with using cloud models for scientific applications~(\autoref{sec:background:cloud}). We then analyse what support representative APIs for shared memory and message passing require from their execution environment~(\autoref{sec:background:code}). Based on this, we develop a list of features necessary to deploy such applications in serverless cloud~(\autoref{sec:background:missing}).

\subsection{Cloud models for scientific applications}
\label{sec:background:cloud}

Cloud platforms for scientific applications such as AWS Batch~\cite{AWSBatchWebsite} and Azure Batch~\cite{AzureBatchWebsite} allocate a dedicated pool of VMs to execute jobs. Each VM has the same size, as determined by the number of vCPUs and memory in GB. For general purpose VMs, memory increases linearly with vCPUs and price per hour~\cite{AzureVMSize}. Jobs in a queue are assigned VMs~\cite{AzureBatchMPI} based on their requested parallelism. For example, MPI jobs specify the number of processes through the command line: \code{mpirun -np <num\_processes>} and are assigned enough VMs such that the sum of vCPUs is greater or equal to the requested processes.

This introduces efficiency challenges due to fragmentation: if a job does not use all of its assigned resources, these resources are wasted. For example, the number of MPI process requested may not be a multiple of the number of vCPUs per VM. In addition, different jobs cannot execute concurrently on the same VM~\cite{AzureBatchLimitation}. In all of these cases, users pay for under-utilised or idle resources, and providers cannot allocate these resources to other jobs in the queue.

Reducing the VM size reduces fragmentation but impacts performance: message passing jobs become less co-located, and shared memory jobs have less available memory. Allocating a mix of different VM sizes in the pool only partially alleviates these problems, because the resource requirements of jobs are unknown ahead of time, making it challenging for providers to provide the right distribution of VM sizes.



In response, cloud platforms have increasingly adopted fine-grained distribution to reduce costs for users, and increase infrastructure utilisation~\cite{BerkeleyViewCloud, BerkeleyViewServerless}. This has culminated in today's serverless cloud offerings, such as AWS Lambda~\cite{AWSLambdaWebsite} and Azure Functions~\cite{AzureFunctionsWebsite}, in which providers take full control of the parallelism and distribution of applications, billing users to a millisecond granularity~\cite{OneStepForwardTwoStepsBack:2019}.

Serverless applications are divided into thousands of small stateless tasks, which can be parallelised and distributed~\cite{AWSLambdaWebsite,AzureFunctionsWebsite,GoogleCloudFunctionsWebsite}. This gives providers the flexibility to allocate resources, and execute functions on those resources, according to bespoke policies. The finer-grained the serverless functions become, the higher the packing density that the provider can achieve and the more control a provider has over the resource allocation to each application.





\subsection{Shared memory and message passing APIs}
\label{sec:background:code}

Although scientific applications cover diverse domains, many employ the same two programming models: shared memory and message passing. Shared memory is used for parallelism within a single machine, \eg using multi-threading libraries such as OpenMP~\cite{OpenMPWebsite}; and message passing is used for parallelism across machines, \eg using MPI~\cite{MPIWebsite} and multi-processing. \autoref{tab:libraries} shows the most starred open-source repositories in several scientific application domains on Github, which all use either or both of these programming models.

OpenMP and MPI, along with other parallel programming models, offer high-level declarative APIs that impose certain features on the underlying execution environment. If we want to deploy such applications with fine-grained distribution in a serverless cloud environment, the cloud platform must offer support to parallelise the computation, and partition, distribute and synchronise private and shared data.

\begin{lstlisting}[float=t,label={lst:sm},style=cpp,language=C++,
caption={Pseudocode for machine learning training using OpenMP's \code{parallel}
abstraction \textnormal{(Within the \code{parallel} block, the OpenMP runtime controls
the degree of parallelism, and ensures access to and synchronisation of the shared
variable \code{weights}.)}}]
int[] weights = initWeights();

for (int i = 0; i < numSteps; i++) {
#pragma omp parallel shared(weights) {
    int threadNum = omp_get_thread_num();
    int nThreads = omp_get_num_threads();
    updateWeights(weights, threadNum, nThreads);
#pragma omp single
    applyWeights(weights);
}
\end{lstlisting}

\autoref{lst:sm} shows a sample OpenMP implementation of stochastic gradient descent~(SGD)~\cite{SGD:2007}, a core algorithm in machine learning training.  The \code{omp parallel} construct requests that a \code{for} loop be executed in parallel with access to a single shared variable, the \code{weights} vector.  The OpenMP runtime has control over the underlying threads and data partitioning between them.  It must ensure read-only access to the shared address space, except for the \code{weights} vector, which receives synchronised writes from multiple threads.

The environment used to execute this code must spawn and execute parallel threads, each with access to a shared address space, and provide synchronisation primitives for accessing shared variables. This is trivial on a single host but becomes difficult if we distribute the computation. In a distributed environment, the shared address space must be made available and synchronised across VMs. Any coordination primitives, such as locks, must also operate in a distributed manner.

As another example, \autoref{lst:mp} shows an SGD implementation based on MPI's \code{MPI\_Allreduce()} operation: concurrent processes execute the same function to transform and aggregate results on multiple VMs. The MPI runtime manages the processes, data transfers and messaging topology: it sums the weights and broadcasts the result to all processes.

\begin{lstlisting}[float=t,label={lst:mp},style=cpp,language=C++,
caption={Pseudocode for machine learning traning using MPI's
\code{MPI\_Allreduce} function \textnormal{(The MPI runtime controls the degree of
parallelism, data partitioning and messaging topology.)}}]
int worldSize, rank;
MPI_Comm_size(MPI_COMM_WORLD, &worldSize);
MPI_Comm_rank(MPI_COMM_WORLD, &rank);
int[] weights = initWeights();

for (int i = 0; i < numSteps; i++) {
  updateWeights(weights, rank, worldSize);
  MPI_Allreduce(MPI_IN_PLACE, weights, nWeights,
      MPI_INT, MPI_SUM, 0, MPI_COMM_WORLD);
  if(rank == 0) applyWeights(weights);
}
\end{lstlisting}

The execution environment for this code must provide a fixed-size pool of processes, each of which maintains in-memory state, and can send/receive messages to/from the others in the group. The processes count cannot be changed throughout the application's lifetime due to the complexity of maintaining consistent in-memory process state, and preserving the rank-based addressing scheme between processes. This makes it challenging to vary the assigned resources.

\subsection{Requirements}
\label{sec:background:missing}

\begin{table}[t]
  \centering
  \footnotesize
  \begin{tabular}{@{}l@{}ccccc@{}}
    \toprule
    \textbf{\makecell{Platform}} &
    \textbf{\makecell{Threads/\\processes}} &
    \textbf{\makecell{Fixed\\parallelism}} &
    \textbf{\makecell{Shared\\addr. space}} &
    \textbf{\makecell{Direct\\comms.}} \\
    \midrule
    AWS Batch~\cite{AWSBatchWebsite} & \cmark & \cmark & \xmark & \cmark \\
    Azure Batch~\cite{AzureBatchWebsite} & \cmark & \cmark & \xmark & \cmark \\
    \midrule
    Azure Dur. Funcs.~\cite{AzureDurableFunctionsWebsite} & \xmark & \cmark & \xmark & \xmark \\
    AWS Step Functions~\cite{AWSStepFunctionsWebsite} & \xmark & \cmark & \xmark & \xmark \\
    AWS Lambda~\cite{AWSLambdaWebsite} & \xmark & \xmark & \xmark & \xmark \\
    Azure Functions~\cite{AzureFunctionsWebsite} & \xmark & \xmark & \xmark & \xmark \\
    \midrule
    Crucial~\cite{Crucial:2019} & \cmark & \xmark & \xmark & \xmark \\
    Faasm~\cite{Faasm:ATC:2020} & \xmark & \xmark & \cmark & \xmark \\
    Faastlane~\cite{Faastlane:2021} & \xmark & \xmark & \cmark & \xmark \\
    Kappa\cite{Kappa:2020} & \cmark & \xmark & \xmark & \cmark \\
    SAND~\cite{SAND:2018} & \xmark & \xmark & \xmark & \cmark \\
    \midrule
    \textbf{\sys} & \cmark & \cmark & \cmark & \cmark \\
    \bottomrule
  \end{tabular}
    \caption{Cloud support for scientific workloads \textnormal{(We differentiate between batch-compute, serverless-compute and academic.)}}
  \label{tab:features}
\end{table}


To perform fine-grained distribution of the code in \autoref{lst:sm} and \ref{lst:mp}, a cloud platform must support the following features: thread/process semantics, fixed parallelism, shared address space and direct communication. \autoref{tab:features} summarises the support for these features in today's scientific cloud platforms, and contrasts them with serverless platforms.



\mypar{Thread/process semantics} When an application uses threads or processes for parallel computation, the execution environment must (i)~fork and join child processes, which duplicate the parent's process state; and (ii)~spawn and join threads, which share the parent's address space.

Existing scientific and serverless cloud platforms support thread/process semantics, but only \emph{within} a given VM or serverless function. Although each VM/function can fork a process/thread, the degree of parallelism is limited by the resources allocated to the host. For example, on AWS~Lambda~\cite{AWSLambdaWebsite}, it is possible to spawn new threads but they run inside the same function, competing for the same resources.  This limits the flexibility afforded to the provider, preventing them from arbitrarily distributing each thread and process on any VM, \eg to achieve optimal bin-packing of multiple tenants.

\mypar{Control of parallelism} Typically, parallel applications use a known number of threads/processes and employ synchronisation primitives (\eg barriers, mutexes and locks) to ensure correct execution. A known parallelism degree allows applications to partition data and computation appropriately.

Existing scientific cloud platforms~\cite{AWSBatchWebsite, AzureBatchWebsite} provide a fixed level of parallelism to each job. They only allocate each VM to one job, resulting in unused resources if the job does not exploit a VM's full parallelism. Serverless platforms, on the other hand, do not guarantee a fixed level of parallelism, instead allocating available resources from the shared infrastructure. While this means that serverless platforms cannot execute scientific workloads, they can perform fine-grained distribution according to their own scheduling policies. In an ideal scientific computing environment, the platform would both guarantee a fixed level of parallelism, while retaining control over its own fine-grained scheduling decisions.


\mypar{Shared address space} As shown in~\autoref{lst:sm}, multi-threaded applications assume that threads share the same virtual address space. To avoid concurrency issues, access to the address space must be coordinated. Standard libraries~\cite{PthreadsManPage, CppThreadsReference} and OSs provide implementations of synchronisation primitives such as mutexes, semaphores and barriers.

As with thread/process semantics, today's cloud-based scientific and serverless platforms both support shared memory within a single \mbox{VM/function}. This means that shared memory parallelism is limited to the scale of a single \mbox{VM/function}, preventing the provider from arbitrarily distributing threads. The ideal scientific computing environment would be able to distribute threads, while maintaining a distributed shared address space across them, as well as providing distributed coordination primitives.


\begin{figure}[t]
  \centering
  \includegraphics[width=.9\linewidth]{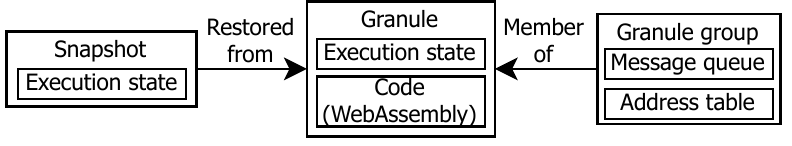}
  \caption{Key abstractions in \sys (\textnormal{\funcs are restored from snapshots, and each \func is a member of a \group).}}
  \label{fig:funcs:func}
\end{figure}

\mypar{Direct communication} Each process in a message-passing application must be able to transfer data to other processes~(see \autoref{lst:mp}). In MPI, this is done based on an address represented as an integer~(rank).

Existing scientific cloud computing platforms support low-latency point-to-point communication for applications, as long as the available parallelism is sufficient to execute a fixed-size pool of threads. These platforms, however, cannot redistribute resources within a running application, as they cannot migrate processes between VMs. Serverless platforms isolate functions even if they belong to the same application, preventing them from obtaining stable identities for communication. An ideal scientific cloud environment would enable long-lived stateful processes with direct communication, yet allow the resources allocated for these processes to be migrated and shared between applications.




%% file: sections/section3.tex

\section{Executing Threads and Processes}
\label{sec:funcs}

\sys is a serverless cloud runtime that supports multi-threading and multi-processing, yet offers fine-grained distribution. It introduces a new parallel computing primitive, the \emph{\func}, with thread and process semantics~(\autoref{sec:funcs:funcs}). \funcs allow \sys to control an application's parallelism and distribution via \emph{\cps}~(\autoref{sec:funcs:points}). At \cps, the \sys runtime interrupts the application to add, remove or migrate \funcs~(\autoref{sec:funcs:migration}). \sys thus efficiently executes parallel applications and gives the cloud provider control over scheduling~(\autoref{sec:funcs:arch}).

\subsection{\func abstraction}
\label{sec:funcs:funcs}

\begin{figure}[t] \centering
  \includegraphics[width=0.9\linewidth]{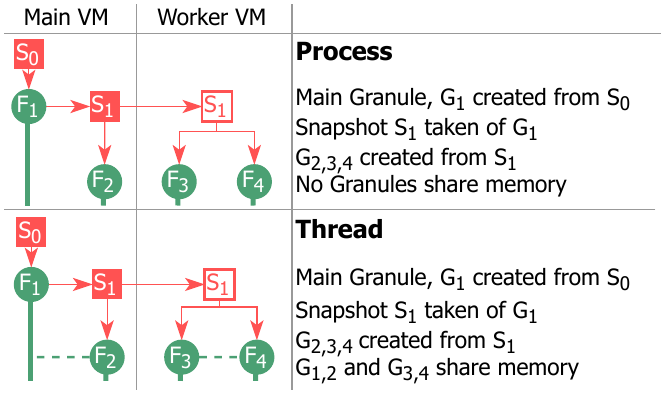}
  \caption{Thread/process semantics with \funcs}
  \label{fig:funcs:thread-proc}
\end{figure}

To support multi-threading and multi-processing, \sys uses \funcs. \funcs can be snapshotted and restored across VMs to support the parent/child semantics of threads and processes. They also share a single distributed address space for shared memory programming~(\autoref{sec:memory}). Finally, they support the direct exchange of messages within \groups for message passing~(\autoref{sec:message}).

\autoref{fig:funcs:func} gives an overview of the key abstractions. Each \func executes application code compiled to WebAssembly~\cite{Wasm:2017}, a binary platform-independent execution format. The use of WebAssembly enforces lightweight memory safety: its isolation mechanism allows \funcs to execute side-by-side in a single instance of the \sys runtime. It also allows for an efficient snapshotting mechanism because the complete execution state of a \func is captured in the single linear memory array of a WebAssembly module.

\sys creates a \func by restoring it from a \emph{snapshot}, which has a copy of a \func's execution state: its linear memory, mutable global variables, a function table and stack pointers. To restore a \func, \sys copies the stack pointer, function table and globals from the snapshot into the \func, and creates a copy-on-write mapping of the \func memory onto the snapshot's linear memory.

\autoref{fig:funcs:thread-proc} shows how \sys uses \funcs and snapshots to replicate process and thread semantics across VMs. Each application has a \emph{base snapshot}, whose linear memory contains the static data of the application. \funcs restored from point-in-time snapshots of their parent \funcs; \funcs with \emph{thread} semantics share a single linear memory mapping with other \funcs on that VM.

\subsection{Intercepting execution with \cps}
\label{sec:funcs:points}

\begin{figure}[t] \centering
    \includegraphics[width=0.8\linewidth]{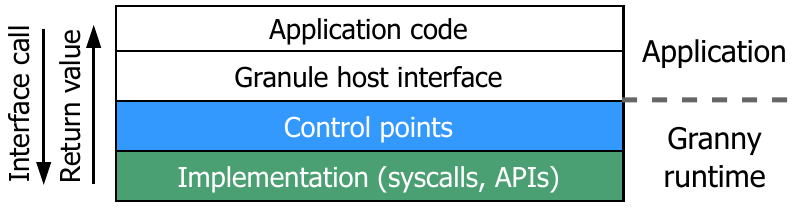}
    \caption{Control points \textnormal{(Control points are triggered when application code calls functions from supported APIs, before the \sys implementation of the given function is executed.)}}
    \label{fig:funcs:points}
\end{figure}

To execute and distribute scientific applications in a cloud environment, \sys must interrupt the application execution periodically: it must (i)~provision new \funcs when the degree of parallelism used by the application changes; (ii)~migrate \funcs as dictated by the scheduler to improve locality and utilisation; (iii)~synchronise shared memory between VMs; and (iv)~deliver messages between \funcs.

As shown in \autoref{fig:funcs:points}, \sys triggers \emph{\cps} when an application invokes certain system calls and parallel programming APIs. \funcs execute application code compiled to WebAssembly, and WebAssembly can pass control to the execution runtime for arbitrary listed functions. \sys uses this approach to transfer control to the runtime on system calls related to thread and process operations, \eg \code{pthread\_create()} and \code{fork()}, as well as functions from OpenMP and MPI APIs, \eg \code{MPI\_Allreduce()}.

When making such function calls, control passes to the \sys runtime at a \cp. Before the runtime executes an API or system call implementation, it may perform one or more of the following actions: (a)~spawn \funcs to execute new logical threads/processes, \eg on \code{fork()}; (b)~await the completion of \funcs to replicate joining a thread or awaiting process completion, \eg on a call to \code{pthread\_join()}; (c)~merge changes to a shared address space using \diffs~(\autoref{sec:memory}), \eg when completing OpenMP \code{parallel} sections; (d)~send/receive messages between the \funcs~(\autoref{sec:message:groups}), \eg due to \code{MPI\_Send()}; and (e)~migrate a  \func~(\autoref{sec:funcs:migration}) to another host, \eg on a call to \code{MPI\_Barrier()}.

\subsection{Migrating \funcs across VMs}
\label{sec:funcs:migration}

\begin{figure}[t] \centering
  \includegraphics[width=.8\linewidth]{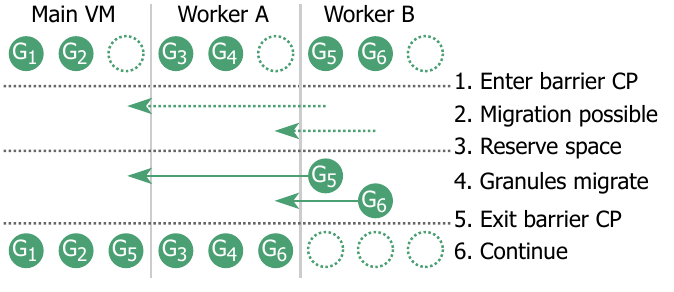}
  \caption{\func migration at barrier \cps}
  \label{fig:funcs:migration}
\end{figure}

Cloud providers must retain control over the scheduling of \funcs on VMs, \eg to increase host utilisation or to co-locate \funcs belonging to the same tenant. \sys achieves this control despite the long-lived execution of \funcs because \funcs can be \emph{migrated} between VMs. Migration decisions are determined by a \emph{scheduling policy}, \eg bin-packing to the fewest VMs, load-balancing across VMs, or exploiting locality for \funcs of a single application.

To simplify the migration process, \func migration may only be carried out at \emph{barrier \cps}. These are \cps that block all \funcs of an application, \eg as triggered by calls to OpenMP's \code{barrier} directive, MPI's \code{MPI\_Barrier()} or \code{MPI\_Allreduce()} functions.

\autoref{fig:funcs:migration} illustrates \func migration. When \funcs reach a barrier \cp, they wait for a notification from the application's main VM. When the main \func reaches the barrier \cp on the main VM, it queries the \emph{scheduler} for migration decisions. The scheduler, periodically and in the background, applies its scheduling policy and decides on function migrations if the current function execution deviates from the desired allocation. It then sends messages to all \sys runtimes on the VMs involved in the migrations.

To migrate a \func, the involved \sys runtimes reserve the necessary resources for the \funcs. If the resources have become unavailable, the migration is aborted. After that, the \funcs to be migrated perform the migration and notify the main VM. Once the main VM has received notifications from all migrated \funcs, it allows the \funcs to exit the barrier \cp.

The actual migration of a \func is performed via the same mechanism that \sys uses to create child processes and threads. The migrating \func takes a snapshot, and sends the snapshot as part of a migration request to the target VM. The target VM creates a new \func with the required semantics, \ie with a new private mapping of the snapshot{}'s linear memory for a process, or sharing a linear memory mapping with existing \funcs for a thread.

\subsection{\sys architecture}
\label{sec:funcs:arch}

\begin{figure}[t]
  \centering
  \includegraphics[width=0.7\linewidth]{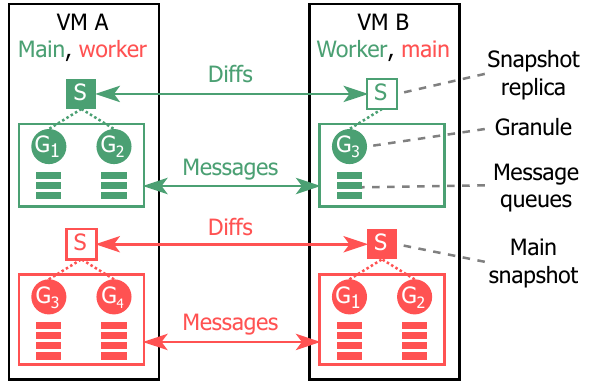}
  \caption{\sys architecture \textnormal{(\sys runtime instances act as either the \emph{main} VM or \emph{worker} VM for each application. They add, remove and migrate \funcs, synchronise replicas of snapshots via \diffs, and asynchronously pass messages between \funcs.)}}
    \label{fig:funcs:overview}
\end{figure}

\autoref{fig:funcs:overview} shows \sys{}'s architecture. A \sys runtime executes on each VM, controlling a variable-sized pool of \funcs, snapshots and message queues. Each \func runs a single thread/process from an application.

\sys uses a distributed shared state \emph{scheduler}: the \sys runtime on each VM has a \emph{local} scheduler, which communicates with the other schedulers in the cluster. The local scheduler allocates up to one \func per CPU core. If the thread or process executing in a given \func requests more parallelism, the local scheduler creates new \funcs on that VM. If that VM has exhausted its CPU cores, the local scheduler chooses another VM, preferably one that already executes \funcs for that application, as it then holds the application code and \func snapshots in memory. If no such VM is available, the scheduler selects the VM with the most available resources. It then transfers the required snapshot to the new VM, and requests that it create and execute the new \func.

Similar to most MPI implementations, \sys currently does not offer fault-tolerance features---when one or more \funcs fail, the whole application fails. Fault tolerance could be added to \sys by exploiting \func snapshots as checkpoints. If a \func fails, the most recent snapshot can be restored, either on the same or on a different VM. Incoming messages to a \func can be persisted by the local \sys runtime until the next snapshot is reached, and replayed after \func failure.


%% file: sections/section4.tex

\section{Shared Memory Programming}
\label{sec:memory}

To execute multi-threaded applications correctly with shared memory, \sys must provide consistency guarantees across \funcs and VMs, alongside suitable synchronisation primitives. \sys makes shared memory consistent by sending \emph{\diffs} between \funcs and VMs that communicate updates to the shared address space~(\autoref{sec:memory:sync}).

Programming models such as OpenMP use \emph{reductions} to aggregate parallel updates to shared variables without the coordination overhead of mutexes. \sys supports reductions using \emph{merge operations}, which allow it to combine multiple \diffs to the same memory region using arithmetic operations~(\autoref{sec:memory:reduction}). Finally, \sys provides custom implementations of coordination primitives including mutexes, barriers and latches~(\autoref{sec:memory:coordination}).

\subsection{Synchronising changes to shared memory}
\label{sec:memory:sync}

By default, multi-threaded applications assume only weak consistency guarantees on the memory shared between threads; stronger consistency is requested explicitly through synchronisation primitives. Assuming code is free from data races, \sys must correctly execute multi-threaded applications: it must ensures that writes to shared memory from a child thread are visible to the parent thread when it joins that child. Changes must be visible to all threads when entering a critical section, or exiting an explicit or implicit barrier~\cite{OpenMPWebsite}.

When \sys needs to execute a child thread, it creates a new \func from a snapshot of the main \func. This snapshot is maintained until all child threads have finished execution, and acts as the \emph{main snapshot} for the shared address space. \sys synchronises all subsequent changes across \funcs via this main snapshot: it receives updates to it from other \funcs and VMs, and uses it to calculate updates to send to remote VMs.

A \func maps its linear memory from the main snapshot if executing on the main VM, or a replica of the main snapshot if executing on a worker VM. The \func then tracks the changes made to the shared address space by application code. \sys write-protects all memory pages of the \func{}'s linear memory using \code{mprotect()}~\cite{MprotectManPage} and handles the page faults caused by application code by marking the page dirty and resetting its read/write permissions.

To send these changes back to the main VM when the \func completes or reaches a barrier, the \func performs a byte-wise comparison of the modified pages with its local copy of the main snapshot. This results in a list of \emph{\diffs} that specify the offset at which the changes occurred and the modified bytes. The main VM receives these \diffs from worker VMs and uses them to update the main snapshot.

The \sys runtime must update the main snapshot replicas on remote VMs, \eg when exiting a barrier. It transmits only the \diffs required to update the remote replicas, and not the whole snapshot. To enable this, the \sys runtime on the main VM keeps track of which bytes have been updated by incoming \diffs, then sends a new set of \diffs with these changes to the worker VMs.

\begin{figure}[t]
  \centering
  \includegraphics[width=\linewidth]{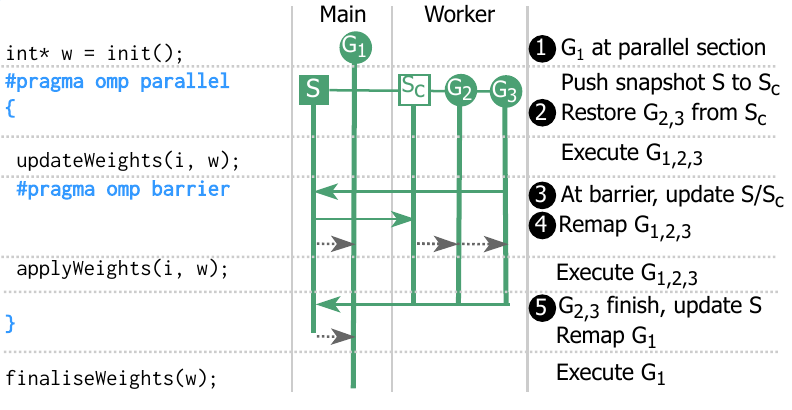}
  \caption{Synchronisation of a shared address space \textnormal{(Distributed \funcs execute an OpenMP \code{parallel} section with a \code{barrier}.)}}
  \label{fig:memory:sync}
\end{figure}

\autoref{fig:memory:sync} gives an example of shared memory synchronisation, which shows how \sys executes an implementation of the SGD example~(\autoref{lst:sm}) using 3~\funcs across 2~VMs. \myc{1} When the main \func enters an OpenMP \code{parallel} section, it triggers a \cp at which \sys creates the main snapshot of the shared address space from the main \func. \myc{2} \sys then creates 2~more \funcs from a replica of this main snapshot on the worker VM, and each \func executes the body of the \code{parallel} section. \myc{3} When all \funcs have reached the \code{barrier}, each creates a list of \diffs that the \sys runtime on the main VM uses to update the main snapshot. \myc{4} On exiting the barrier, the main VM's \sys runtime sends another list of \diffs to update the snapshot replica on the worker VM. All \funcs then remap their own linear memory to the local copy of the snapshot and continue execution. \myc{5} At the end of the \code{parallel} section, the parent \func joins the child \funcs, and again \sys uses the \diffs from each \func to update the main snapshot. Finally, the main \func remaps its memory from the snapshot.

\subsection{Supporting reductions on shared variables}
\label{sec:memory:reduction}

\begin{figure}[t] \centering
    \includegraphics[width=0.9\linewidth]{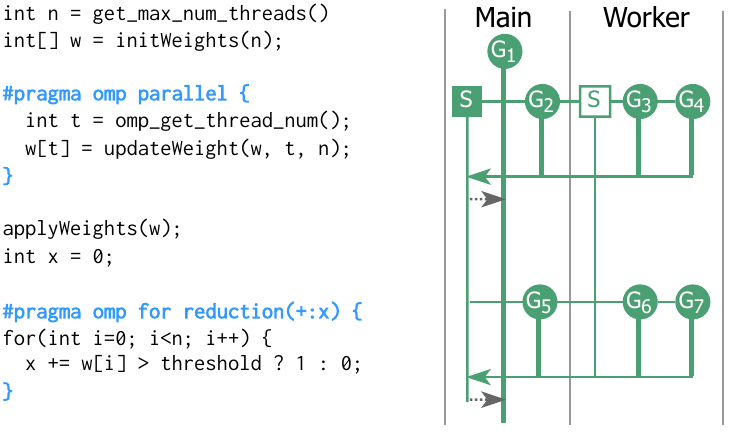}
    \caption{Reductions in OpenMP \textnormal{(\funcs on 2~VMs make non-conflicting updates in a \code{parallel} section, and perform a reduction.)}}
    \label{fig:memory:reduction}
\end{figure}

Declarative multi-threading frameworks such as OpenMP allow multiple threads to aggregate changes to shared variables using \emph{reductions}~\cite{OpenMPReductions}. A reduction spawns multiple threads that update one or more shared variables in parallel, then aggregates those updates once all threads have completed execution, \eg via a summation. By deferring the aggregation of concurrent updates to shared variables until threads have completed, we can avoid the overhead that would otherwise be incurred from synchronising those updates using a mutex. \sys distributes multi-threaded applications, hence reducing synchronisation between threads via reductions reduces cross-VM coordination overheads.

To distribute reductions in \sys, the runtime on each VM performs the reduction operation locally for the \funcs executing on that VM, then transmits its updates back to the main VM as a \diff. For each \diff sent back to the main VM, \sys specifies a \emph{merge operation}, with the arithmetic operation that should be used to apply that \diff to the main copy of the shared variable.

\autoref{fig:memory:reduction} shows OpenMP code for a \code{parallel} section that performs disjoint updates to a shared vector and a \code{reduction} section to update a shared variable via a summation. \sys spawns 3~child \funcs when the main \func reaches the parallel section, creating the main snapshot on the main VM and a replica on the worker VM. Each \func maps its linear memory from its local copy of the snapshot.

In the first parallel section, each \func updates its value in the \code{w} vector. The resulting \diffs can be written directly to the main snapshot without a merge operation. In the \code{reduction} section, each thread updates their local copy of the variable~\code{x}, generating a \diff on the same memory region. Since the \code{reduction} specifies a summation over~\code{x}, \sys combines these \diffs in the main snapshot using a \code{sum}.

\autoref{tab:merge} lists the merge operations supported by \sys. They include simple arithmetic operations of summation, subtraction, multiplication and division, as commonly found in parallel reductions. The operations involve four values: $A_0$, the starting value in the main snapshot; $B_0$, the value held in the copy of the snapshot on the remote VM; $B_1$, the updated value after the thread has executed on the remote VM; and $A_1$, the value written to the main snapshot by the operation.

\begin{table}[t]
    \centering
    \footnotesize
    \begin{tabular}{lll}
        \toprule
        \textbf{Merge operation}&
        \textbf{Formula}&
        \textbf{Data types}\\
        \midrule
        sum & $A_1 = A_0 + (B_1 - B_0)$ & All numeric \\
        subtract & $A_1 = A_0 - (B_0 - B_1)$ & All numeric \\
        multiply & $A_1 = A_0\, *\, (B_1 \, /\,  B_0)$ & All numeric \\
        divide & $A_1 = A_0\, /\, (B_0 \, / \, B_1)$ & All numeric \\
      overwrite & $A_1 = B_1$ & Arbitrary bytes \\
      \bottomrule
    \end{tabular}
    \caption{Merge operations supported by \sys. \textnormal{(\sys overwrites
    the original value~$A_0$ in the main snapshot with value~$A_1$. $B_0$ is the value seen in the snapshot on the remote VM before the \func executed, and $B_1$ is the value after execution.)}}
    \label{tab:merge}
\end{table}

\subsection{Synchronisation primitives}
\label{sec:memory:coordination}

In addition to providing shared memory and reduction operations, \sys must support the synchronisation primitives in multi-threaded code that control concurrent access to shared data. \sys offers the following primitives:

\mypar{Mutexes} A mutex guarantees that only one \func can access data at a given time. In \sys, application code that acquires a mutex triggers a \cp, and the associated \func requests a lock on the mutex from the \sys runtime. When locking the mutex, the \sys runtime returns the \diffs to update that \func{}'s local copy of the shared memory snapshot. This way, the \func holding the mutex is guaranteed to observe the updates of other \funcs that have also held it; when releasing the mutex, the \func returns its own set of \diffs to the \sys runtime.

\mypar{Atomic operations} Atomic arithmetic operations do not guarantee consistency, only atomicity. To perform such operations, each \func acquires a VM-local mutex to avoid data races on the local copy of the shared memory. \sys then uses a merge operation corresponding to the arithmetic operation to merge the resulting \diffs.

\mypar{Barriers} A barrier is either implicit or explicit: an implicit barrier is introduced by a \code{parallel} section; an explicit barrier is added manually. Barriers require that all \funcs block until they have completed the barrier. Afterwards, all \funcs must observe a consistent view of the shared memory. On entering a barrier, \funcs send their \diffs to the main VM and block. After all threads have completed, the main VM sends the aggregated \diffs to all \funcs, which unblock.

\mypar{Latches} A latch allows \funcs to decrement a counter and/or wait for it to reach zero. Latches are used implicitly in \code{nowait} OpenMP operations: the main \func blocks until all child \funcs have reached the latch. \funcs can make non-blocking requests to the main VM to decrement a latch, or blocking requests to wait for the latch to reach zero.


%% file: sections/section5.tex

\section{Message Passing}
\label{sec:message}

To support multi-process applications with message passing, \sys must associate \funcs with long-lived identities for communication. \sys organises \funcs into \emph{\groups}, in which each \func is assigned an index. It can then asynchronously send and receive messages to and from other \funcs in the group by referring to that index~(\autoref{sec:message:groups}).

When migrating \funcs between VMs, each \sys runtime instance updates its local \group metadata to ensure consistent message delivery independent of each \func's placement~(\autoref{sec:message:migration}). For efficient collective communication, \sys provides \group{}-aware implementations of operations such as all-reduce and broadcast, maximising fast intra-VM messaging via in-memory data structures~(\autoref{sec:message:collective}).

\subsection{Communication in \groups}
\label{sec:message:groups}

Each \func that executes a process in a multi-process application may need to send messages directly to another \func, \eg to fulfil API calls in a message-passing framework such as MPI. To enable such message passing efficiently, \sys allows asynchronous messaging between \funcs within a \group. The asynchronous nature of \sys's message passing avoids blocking a sender \func while the receiver \func is being migrated or initialised.

By default, all \funcs that execute an application are in the same \group. \sys may create new \groups on \cps, \eg as triggered by \code{MPI\_Comm\_create()}, which allows the application to control communication groups. A \group assigns each \func an \emph{index}, which \sys uses as an address for that \func in its implementation of message passing APIs. Each \sys runtime that executes a \func holds a replica of the \group metadata with an \emph{address table} that maps the indexes of \funcs in the \group to the VM on which they have been scheduled. For each \func of the group executed on a VM, the \sys runtime has a set of queues to buffer messages sent to that index.

When a \func triggers a \cp that requires sending a message, \eg \code{MPI\_Send()}, the \sys runtime on the VM of the sending \func looks up the recipient index in the address table for that group. If the recipient is on the same VM, \sys directly enqueues the message on the relevant in-memory queue. This results in low-latency intra-VM messaging compared to using the local loopback network interface or inter-process communication~(IPC). If the recipient is on another VM, \sys sends the message to the runtime on that VM, where it is enqueued.

\subsection{Groups when migrating \funcs}
\label{sec:message:migration}

\begin{figure}[t] \centering
    \includegraphics[width=\linewidth]{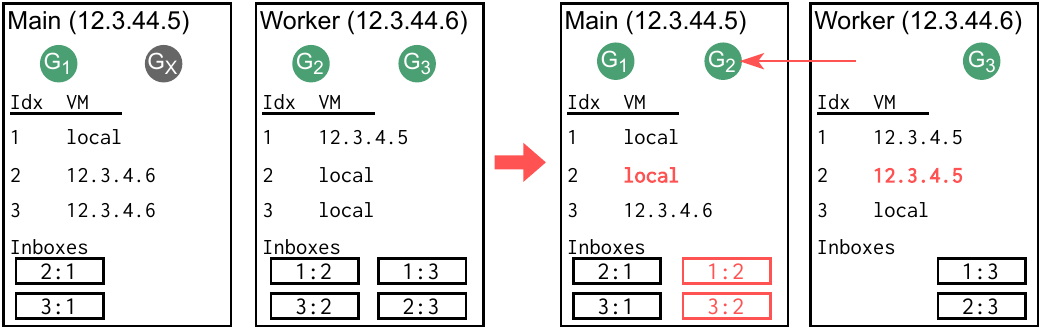}
    \caption{Preserving a \group while migrating \textnormal{(One \func from the \group is migrated from a worker VM to the main VM, and \sys updates the metadata and queues.)}}
    \label{fig:message:migration}
\end{figure}

When migrating \funcs across VMs~(\autoref{sec:funcs:migration}), the \sys runtimes must also update the metadata and queues associated with \groups to which the migrating \funcs belong.

\autoref{fig:message:migration} shows how \sys updates a \group during migration. Initially, the main VM executes one \func from the group alongside a \func from another application; the worker VM executes two other \funcs from the group. When the \func from the other application completes, it frees up resources on the main VM; when the \funcs reaches a barrier \cp, \sys migrates one of the \funcs from the worker VM to the main VM. Before completing the migration, each \sys runtime updates its address table and creates or deletes queues to accommodate the new or departing \func, respectively.

To avoid issues with message delivery arising from migrated \funcs while messages are in-flight or queued, \sys only inserts barrier \cps on message passing operations where \funcs in worker VMs are waiting for one \func in the main VM, \eg MPI's \code{MPI\_Barrier()}, \code{MPI\_Allreduce()} and \code{MPI\_Gather()}. This way, \sys guarantees that there are no messages in-transit.

\subsection{Collective communication}
\label{sec:message:collective}

\begin{figure}[t] \centering
    \includegraphics[width=\linewidth]{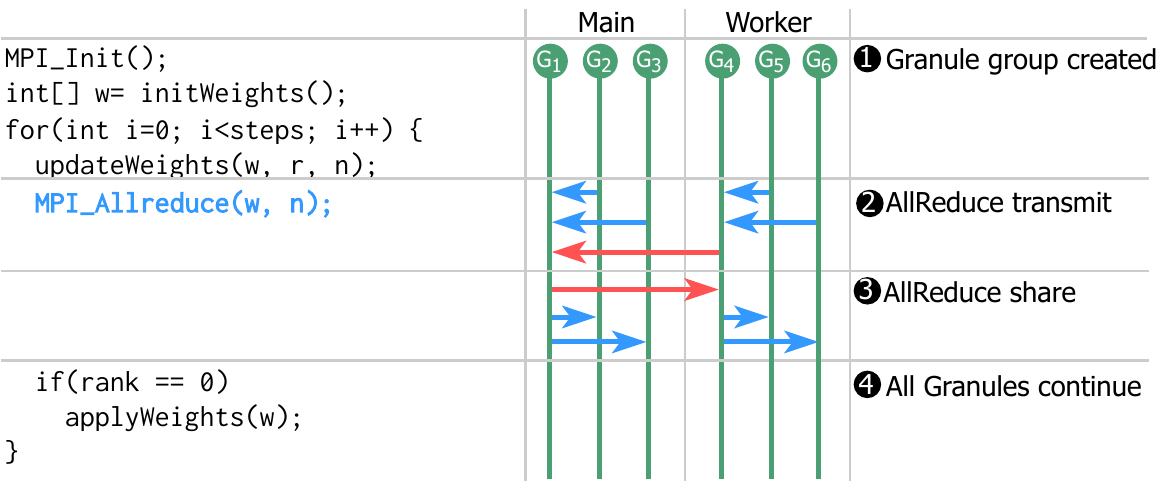}
    \caption{MPI collective communication \textnormal{(A VM-leader on each VM sends/receives intra-VM messages~(blue) and sends/receives cross-VM message~(red) to/from other VM-leaders.)}}
    \label{fig:message:code}
\end{figure}

In addition to simple point-to-point messaging, most message passing applications make use of \emph{collective communication} operations, such as all-reduce and broadcast. These operations are widely used in distributed ML training through specialised libraries~\cite{Hoplite:2021, GlooGithub, OptimalAllReduce:2009}. \sys uses custom \group{}-aware implementations of these operations, which minimise latency by exploiting knowledge of \func placement to maximise intra-VM messaging.

\autoref{fig:message:code} shows the message passing performed by \sys when application code makes a call to \code{MPI\_Allreduce()}. \myc{1}~When \sys creates a \group, it selects one \func on each VM to be the \emph{VM-leader} for that VM. Any messages that need to be sent to \funcs on other VMs are sent by all \funcs to their VM-leader, which batches the messages into single cross-VM requests. All-reduce takes place in two steps: \myc{2} an initial reduce in which results from all \funcs on each VM are sent to the main VM via their VM-leader; and \myc{3} a broadcast of the final result to all \funcs, which is delivered via their VM-leader.

\sys{}'s implementation of collective communication operations reduces the cross-VM messages to one per remote VM involved in each step. It then uses fast in-memory queues for the \func to VM-leader communication. This approach reduces latency (\autoref{subsec:eval:message-passing}) and bandwidth usage, and enables pipelining: after a \func has asynchronously messaged its local leader, it can continue execution.


%% file: sections/evaluation.tex

\section{Evaluation}
\label{sec:evaluation}

Our evaluation answers the following questions: (i)~what are the benefits of using \sys{} to run scientific applications on shared VMs?~(\autoref{subsec:eval:cluster}); (ii)~what is the impact of the cluster size on \sys{}?~(\autoref{subsec:eval:scaling}); (iii)~what is the performance overhead of executing shared memory applications using \funcs{}?~(\autoref{subsec:eval:shared-memory}); (iv)~what is the performance overhead of executing message passing applications using \funcs{}?~(\autoref{subsec:eval:message-passing}); and (v)~what is the performance overhead when migrating \funcs at runtime?~(\autoref{subsec:eval:migration})

\subsection{Experimental set-up}

\mypar{Implementation} \sys is written in 24,000~lines of C++20, compiled using clang-13, and available as open-source at: \emph{removed for anonymity}. Deployed applications and all transitive dependencies, \eg \emph{libc}, are compiled to WebAssembly~\cite{Wasm:2017} using clang-13~\cite{ClangWebsite}, as part of the \sys CPP toolchain, also available as open-source at: \emph{removed for anonymity}. The batch scheduler is written in 1000~lines of Python, and is available at: \emph{removed for anonymity}.

\mypar{Test-bed} We deploy \sys and OpenMPI on a Kubernetes cluster~\cite{KubernetesWebsite} on Azure~\cite{AzureWebsite}. The cluster has 32 \emph{Standard\_D8\_v5} VMs~\cite{AzureVMSize} with 8~vCPU cores and 32\unit{GB} of memory. Native OpenMP applications execute using the underlying VMs in. We implement a batch scheduler that monitors cluster resources and executes jobs as soon as there are enough free resources (in terms of vCPUs).

\mypar{Workloads} We evaluate \sys with scientific applications written using OpenMP~\cite{OpenMPWebsite} for shared memory, and MPI~\cite{MPIWebsite} for message passing. We compare these applications running on \sys against native implementations running directly on VMs. All OpenMP code is compiled using clang-13~(OpenMP~v4.5)~\cite{OpenMPClang}, and we use OpenMPI~v4.1~\cite{OpenMPIWebsite}. \sys is deployed using Kubernetes~(K8s)~\cite{KubernetesWebsite}.


\subsection{Efficiency of running scientific applications}
\label{subsec:eval:cluster}

\begin{figure*}
    \centering
    \begin{subfigure}[b]{0.93\textwidth}
        \includegraphics[width=\linewidth]{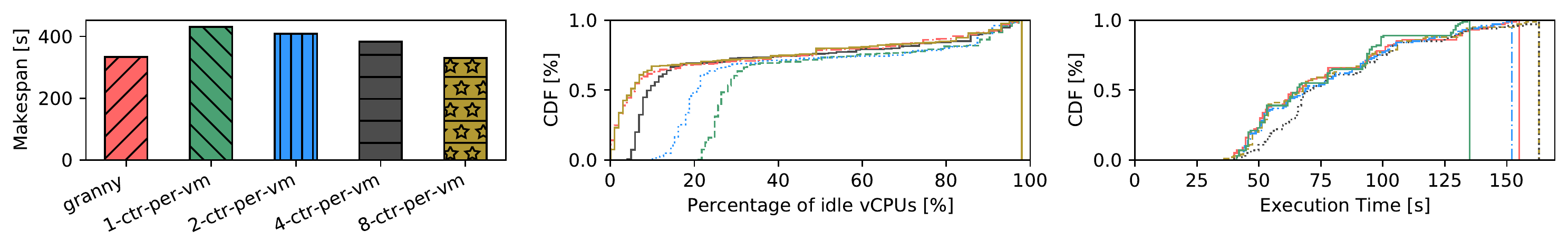}
        \caption{Execution of 100 MPI jobs \textnormal{(Requested parallelism distributed uniformly from 4--16~vCPUs/MPI processes)}}
        \label{fig:eval:cluster-saturation-mpi}
    \end{subfigure}

    \begin{subfigure}[b]{0.93\textwidth}
        \includegraphics[width=\linewidth]{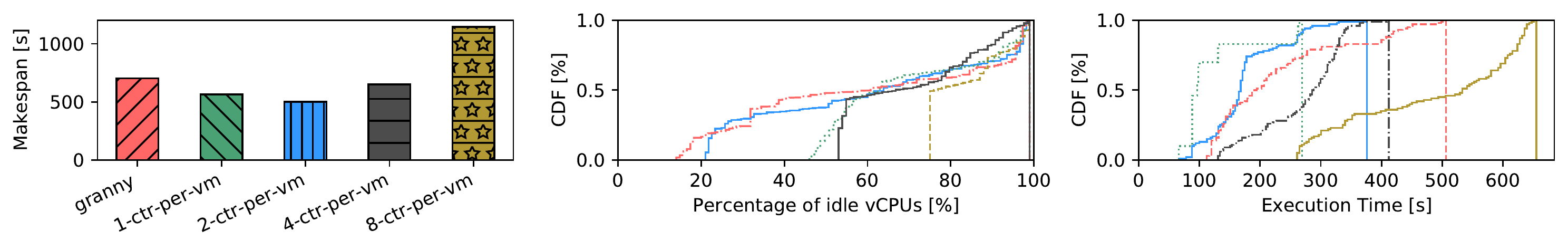}
        \caption{Execution of 100 OpenMP jobs \textnormal{(Requested parallelism distributed uniformly from 2--8 vCPUs/OpenMP threads)}}
        \label{fig:eval:cluster-saturation-omp}
    \end{subfigure}

    \caption{Execution breakdown of running a trace with 100 jobs \textnormal{(We report the total time elapsed, \ie makespan; the CDF of idle vCPUs; and the CDF of job execution times.)}}\label{fig:eval:cluster-saturation}
\end{figure*}

We explore the efficiency and performance impact of using \sys{} to execute applications on a shared VM cluster. As a workload, we generate a trace of 100~jobs. Each job executes a scientific application with a different level of parallelism specified by MPI's world size (as indicated in the \code{mpirun} command) or by the \code{OMP\_NUM\_THREADS} environment variable.

We generate two traces: one with message passing applications (\code{mpi}), and one with shared memory applications (\code{omp}). For the message passing applications, we use LAMMPS~\cite{LAMMPS:1995,LAMMPSWebsite}, a popular molecular dynamics simulator written in C++ using MPI. We pick the Lennard-Jones~(LJ) atomic fluid simulation with 4~million atoms, as it is one of the standard benchmarking problems in the LAMMPS suite~\cite{LAMMPSBenchmarks}. For the shared memory application, we run a dense matrix multiplication~(DGEMM), part of the ParRes kernels~\cite{ParResKernelsGithub}.

As baselines, we want to compare to different VM sizes, exhaustively exploring the fragmentation space: larger VM sizes (\ie higher vCPU counts and larger memory amounts) yield higher per-job performance but result in more idle vCPUs; smaller VM sizes lead to lower per-job performance but fewer idle vCPUs. To avoid managing different VM pools with different VM sizes, we deploy a single VM pool (32~\emph{Standard\_D8\_v5} VMs) and emulate smaller VMs by using 1, 2, 4, or 8~containers, enforcing an even vCPU/memory split using K8s resource limits~\cite{KubernetesResourceLimits}. Different jobs execute in different containers: MPI jobs take up $\lceil \frac{\text{MPI world size}}{\text{vCPUs per ctr}} \rceil$ containers; OpenMP jobs use 1~container, overcommitting vCPUs to threads if \code{OMP\_NUM\_THREADS} > vCPUs per container.



We configure the batch scheduler to schedule jobs in sequence, as soon as there are sufficient vCPUs. This means that the scheduling granularity becomes important: if the cluster is fragmented, jobs will wait for longer in the queue, increasing makespan; over-fragmenting jobs to pack them tightly will increase their execution time, also increasing makespan.


\autoref{fig:eval:cluster-saturation} shows the results. For each job trace (\code{mpi} in \autoref{fig:eval:cluster-saturation-mpi} and \code{omp} in \autoref{fig:eval:cluster-saturation-omp}), we report the total time to execute the 100~jobs (\ie makespan), the CDF of the percentage of idle vCPUs in the cluster, and the CDF of job execution times.

In terms of makespan, MPI jobs in \sys have a 13\%--23\% lower makespan compared to all fixed-sized baselines except for \code{8-ctr-per-vm}. \code{8-ctr-per-vm} is equivalent to running each MPI process in a separate container. Given the MPI's message passing nature, this approach is acceptable and performs on-par with \sys. For OpenMP jobs, however, using 1~container per vCPU reduces performance: \sys's makespan is 38\% lower than \code{8-ctr-per-vm}'s. However, \sys has between 8\% and 25\% higher makespan than the other baselines. This is due to the overhead of \sys{}'s shared memory implementation (see~\autoref{subsec:eval:shared-memory}). In summary, \sys achieves lower makespan than the native baselines, reducing both user costs and provider under-utilisation.



To understand why \sys reduces the makespan, we analyse the CDF of the percentage of idle vCPUs, and the CDF of the job execution time. For MPI jobs, half of time, \sys has at most 5\% of idle vCPUs, whereas the native baselines leave 10\%--30\% of vCPUs idle (with the exception of \code{8-ctr-per-vm}). This shows that \sys better utilised the available vCPUs by packing and distributing jobs at a finer granularity using \funcs. For OpenMP, 100 jobs are not enough to saturate the cluster, specially for baselines that overcommit. Per-job execution time in \sys is on-par with all baselines for the MPI jobs, with the exception of the last 15\% of jobs -- these jobs take longer to execute because they are over-fragmented. For OpenMP jobs, \sys's execution time becomes worse than the baselines due to the overhead of shared memory executions. In summary, \sys achieves lower makespan by allocating resources at finer-grained granularity, paying the price of over-fragmenting some jobs, making them run for longer.


\subsection{Scalability}
\label{subsec:eval:scaling}

\begin{figure}[t]
    \centering
    \begin{subfigure}[b]{.46\linewidth}
        \includegraphics[width=1.05\linewidth]{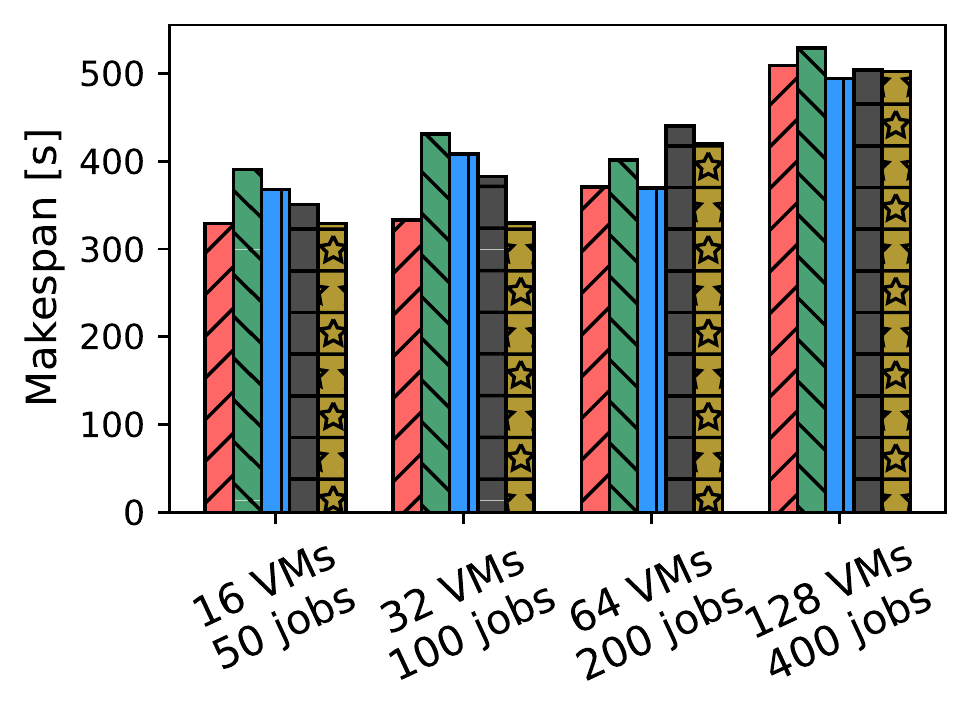}
        \caption{Makespan}
        \label{fig:eval:cluster-scalability:makespan}
    \end{subfigure}
    \begin{subfigure}[b]{.46\linewidth}
        \includegraphics[width=1.05\linewidth]{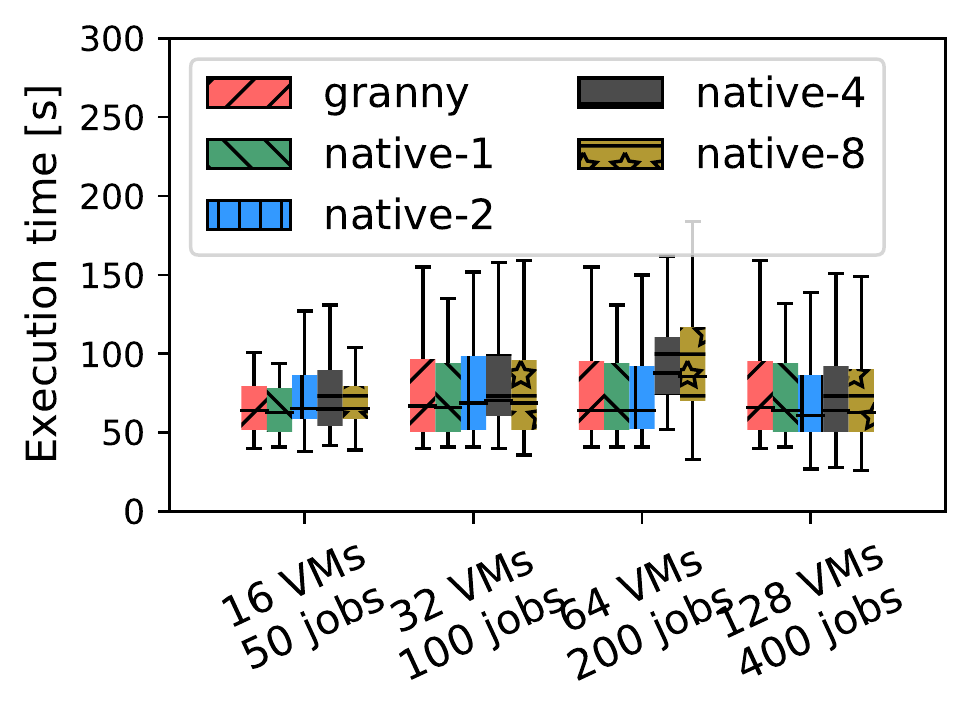}
        \caption{Job execution time}
        \label{fig:eval:cluster-scalability:exec-time}
    \end{subfigure}

    \caption{Scaling the VM count \textnormal{(We report the makespan to execute a batch of jobs, which increases linearly with cluster size, and the distribution of execution times.)}}
    \label{fig:eval:cluster-scalability}
\end{figure}

We explore \sys's scalability with respect to the number of VMs in the cluster. As a workload, we use traces of 50, 100, 200 and 400 MPI jobs in a cluster with 16, 32, 64 and 128 \emph{Standard\_D8\_v5} VMs, respectively. We measure the makespan and the average job execution time. By increasing the number of tasks with the cluster size, we expect the makespan to stay constant. We use the same batch scheduler configuration and baselines from \autoref{subsec:eval:cluster}.

In \autoref{fig:eval:cluster-scalability}, we report, for each cluster size, the make\-span~(\autoref{fig:eval:cluster-scalability:makespan}) and the distribution of execution times with a box plot that includes the median, boxes for the 25\textsuperscript{th} and 75\textsuperscript{th} percentiles, and whiskers extending 1.5$\times$ the inter-quartile range~(\autoref{fig:eval:cluster-scalability:exec-time}).

\sys achieves a 7\%--16\%, 13\%--23\% and 10\%--20\% lower makespan for 16, 32 and 64 VMs respectively, because it manages to utilise the cluster resources more efficiently. Its makespan is on par with \code{native-8}, as explained in \autoref{subsec:eval:cluster}. Note that the makespan values for 16--64 VMs vary within 5\%--10\% of each other, which is caused by the different job sizes in each trace. For 128 VMs, however, the performance of all deployments degrades due to the implementation of our centralised batch scheduler, which becomes a bottleneck. \sys's makespan also is 5\% higher than the baselines because \sys{}'s current implementation centrally manages all registered VMs and their resources.

To explore this performance degradation further, \autoref{fig:eval:cluster-scalability:exec-time} shows the distribution of execution times for each baseline and cluster size using a boxplot. Each job in the trace has a different level of parallelism, and longer traces have more jobs, introducing randomness. In spite of this randomness, \autoref{fig:eval:cluster-scalability:exec-time} shows that the 25\textsuperscript{th}, median, and 75\textsuperscript{th} percentiles are very similar across baselines, and cluster sizes. The whiskers are more variable, as they account for the tails of the distribution. These results confirm our hypothesis that the performance degradation in \autoref{fig:eval:cluster-scalability:makespan} is due to limitations of \sys's cluster management and batch scheduler implementation, rather than \funcs. These limitations can be overcome with additional engineering work or by deploying \sys across multiple 64~VM clusters, and load-balancing requests across them.



\subsection{Shared memory performance}
\label{subsec:eval:shared-memory}

\begin{figure}
    \centering
    \includegraphics[width=\linewidth]{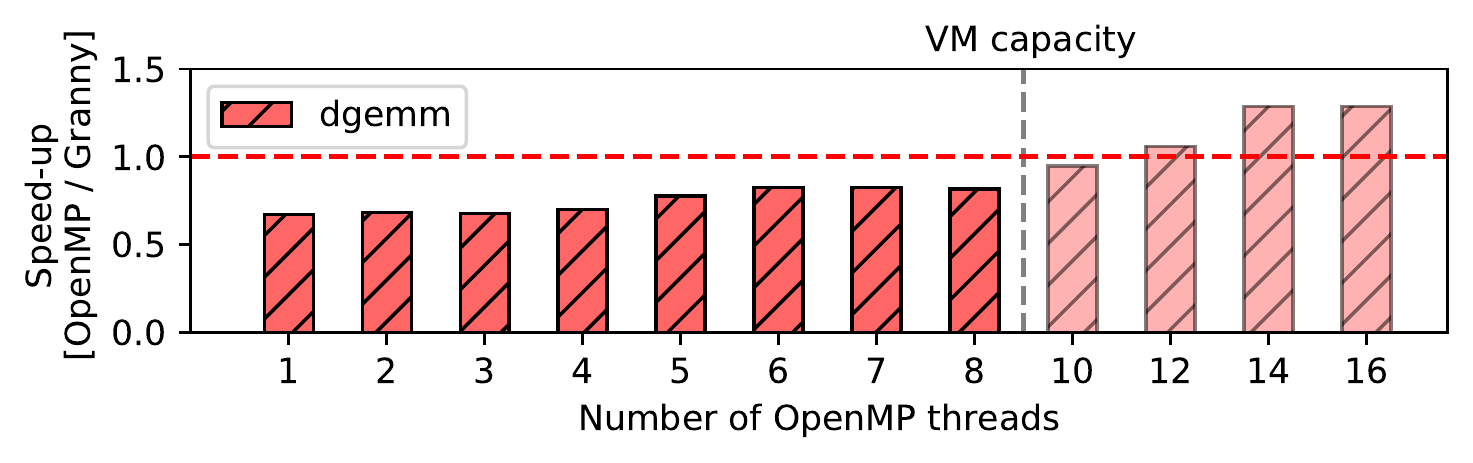}
    \caption{Speed-up executing shared memory applications \textnormal{(When the number of threads spans multiple VMs, we compare \sys's time with OpenMP's execution time with 8 threads.)}}
\label{fig:eval:shared-memory}
\end{figure}

Next we investigate the performance overhead introduced by \sys when executing shared memory applications using OpenMP. We use the same OpenMP application from \autoref{subsec:eval:cluster}: a dense parallel matrix multiplication~(DGEMM) from ParResKernels~\cite{ParResKernelsGithub}. DGEMM is computationally-intensive, with some light use of \sys's shared memory synchronisation and coordination primitives.


\autoref{fig:eval:shared-memory} shows the speed-up achieved by \sys, computed as the ratio between native OpenMP and \sys's execution time. We execute both baselines with different numbers of OpenMP threads: native OpenMP cannot be scaled out, limiting its parallelism to that that of 1~VM (\ie 8 vCPU cores, with each thread pinned to one core); \sys can scale out shared memory applications using \funcs. For distributed execution (\ie with thread counts greater than 8; faded bars in \autoref{fig:eval:shared-memory}), we measure the speed-up as the native execution time with 8 OpenMP threads divided by \sys's execution with the higher thread count.

For DGEMM, \sys is 20\%--30\% slower than native OpenMP in a single VM, due to the overhead of performing floating-point arithmetic in WebAssembly~\cite{NotSoFast:2019}. When scaling out to another VM, \sys achieves the optimal native performance in one VM with 50\% more threads. \sys achieves a 25\% speed-up (over native OpenMP with 8 threads) when executing with twice as many threads. In summary, \sys distributes shared memory applications with \funcs, surpassing the performance of a single VM deployment.




\subsection{Message passing performance}
\label{subsec:eval:message-passing}

\begin{figure}
\centering
    \begin{subfigure}[b]{\linewidth}
        \includegraphics[width=\linewidth]{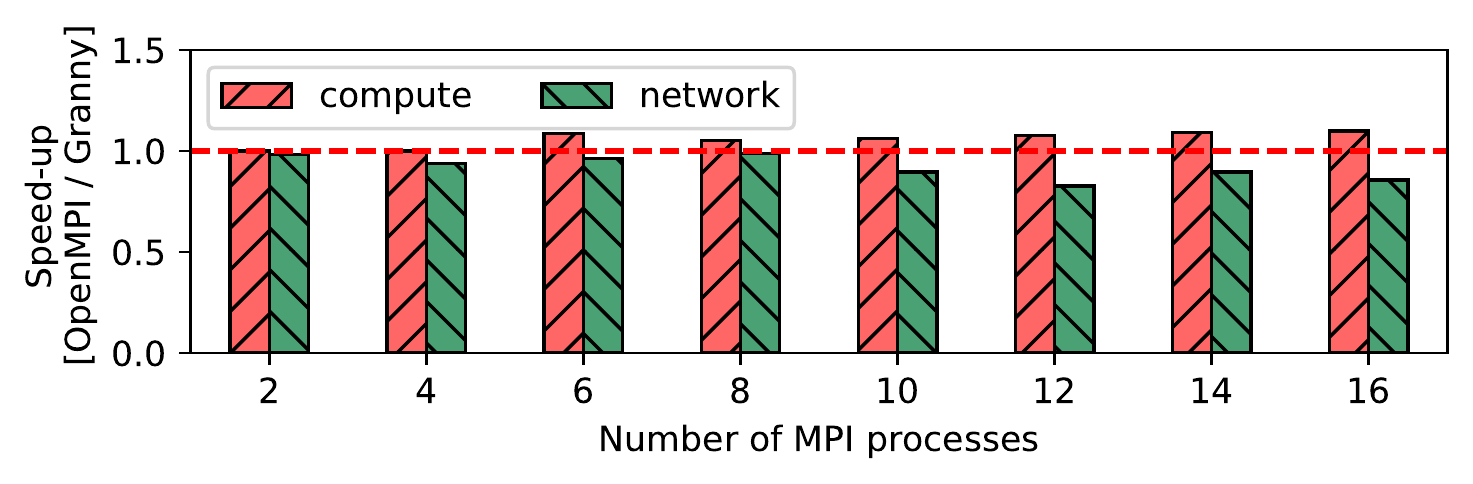}
        \caption{LAMMPS}
        \label{fig:eval:mpi-slowdown:lammps}
    \end{subfigure}

    \begin{subfigure}[b]{\linewidth}
        \includegraphics[width=\linewidth]{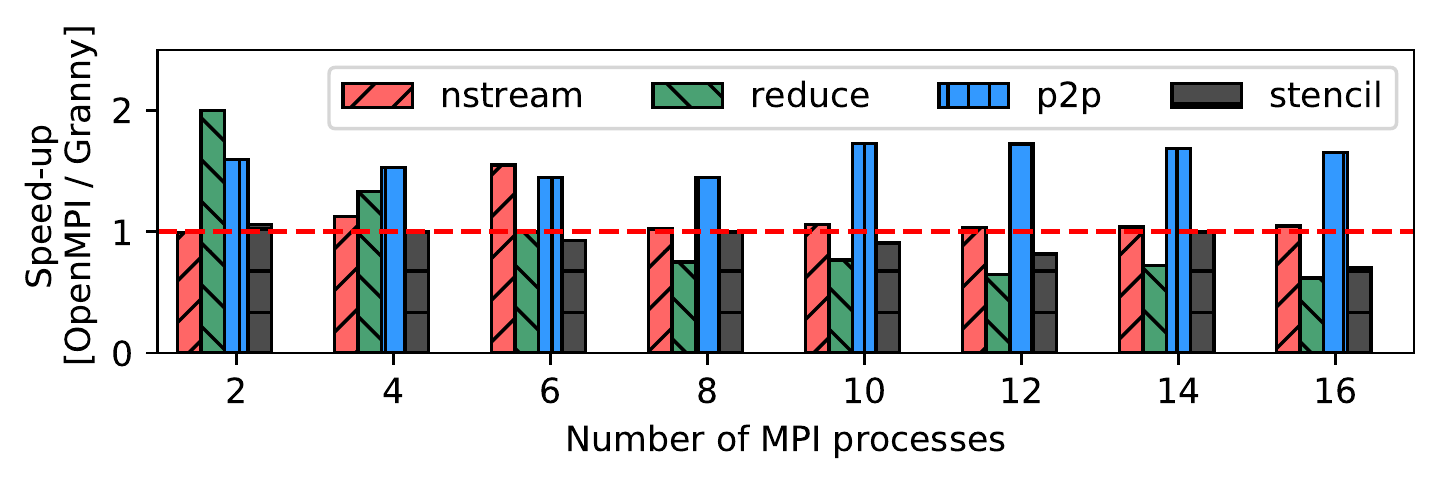}
        \caption{ParRes Kernels}
        \label{fig:eval:mpi-slowdown:kernels}
    \end{subfigure}

    \caption{Speed-up executing message passing applications}
    \label{fig:eval:mpi-slowdown}
\end{figure}

This experiment explores \sys{}'s performance overhead when executing message passing applications using MPI. We run the same MPI application as in \autoref{subsec:eval:cluster}: the LAMMPS simulator running the Lennard-Jones~(LJ) benchmark with 4~million atoms. To stress \sys{}'s communication layer, we update LAMMPS' \emph{controller} example~\cite{LAMMPSController} and increase the synchronisation steps, resulting in three orders of magnitude more cross-VM messaging. We refer to the LJ benchmark as \emph{compute-bound}, and the modified \emph{controller} one as \emph{network-bound}. We also run a subset of the ParRes kernels~\cite{ParResKernelsGithub} to evaluate specific parts of \sys{}'s implementation.


\autoref{fig:eval:mpi-slowdown} shows the speed-up that \sys achieves, computed as the ratio between OpenMPI's and \sys's execution times. We execute the two LAMMPS simulations~(\autoref{fig:eval:mpi-slowdown:lammps}) and the ParRes kernels~(\autoref{fig:eval:mpi-slowdown:kernels}) with different levels of parallelism.

For LAMMPS' compute-bound benchmark, \sys achieves a 5\%--10\% speed-up over OpenMPI. \sys slightly outperforms OpenMPI due to the faster intra-VM messaging using shared memory and the locality-aware collective communication implementation. For the network-bound benchmark, \sys's execution time is up to 15\% higher than OpenMPI's. This slow-down is due to an additional level of indirection in \sys{}'s transport layer to support concurrent applications, which becomes a bottleneck with higher message throughputs.


To analyse the performance of different parts of \sys's message passing implementation further, we execute ParRes kernels~\cite{ParResKernelsGithub} for distributed computation. For the point-to-point messaging kernel (\code{p2p}), \sys achieves a 50\%--70\% speed-up over OpenMPI because most messages are intra-VM, and \sys can use in-memory queues. The \code{nstream} kernel updates an array in streaming fashion and synchronises with a barrier, and \sys matches the OpenMPI performance. For the reduction kernel (\code{reduce}), \sys is between 25\% faster and 25\% slower in a single VM, and 25\%--50\% slower when distributing MPI processes over 2~VMs. \sys performs worse than OpenMPI when there are more cross-VM messages, as previously discussed. Finally, for the \code{stencil} kernel, \sys is up to 30\% slower because cross-VM messages dominate execution.

In summary, \sys's message passing performance is comparable to that of OpenMPI. \sys performs best for intra-VM messages, as it can use its in-memory queues, and worst for cross-VM messages, which add extra overhead.


\subsection{Migration of \funcs}
\label{subsec:eval:migration}

\begin{figure}[t]
    \centering
    \begin{subfigure}[b]{.48\linewidth}
        \includegraphics[width=\linewidth]{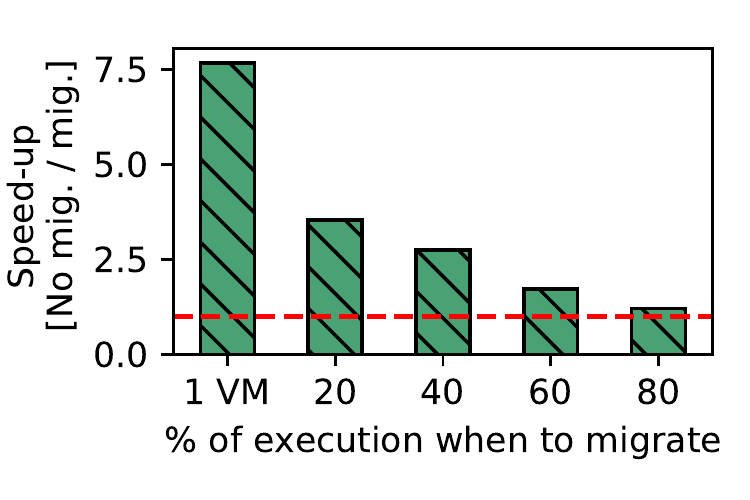}
        \caption{All-to-all}
        \label{fig:eval:migration:all-to-all}
    \end{subfigure}
    \begin{subfigure}[b]{.48\linewidth}
        \includegraphics[width=\linewidth]{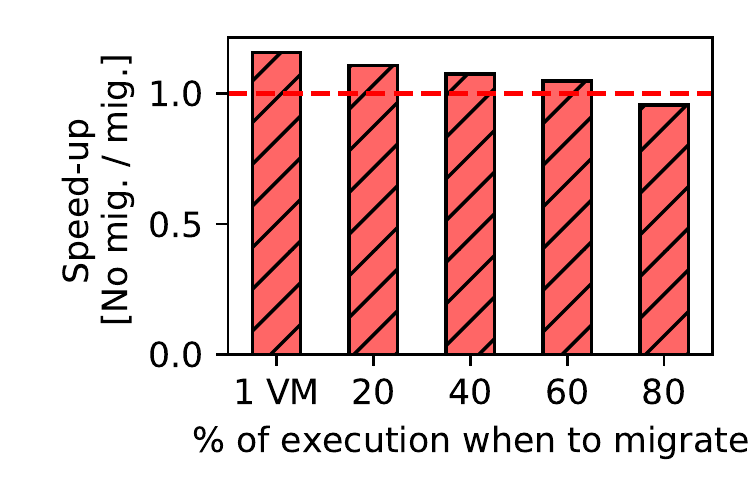}
        \caption{LAMMPS}
        \label{fig:eval:migration:lammps}
    \end{subfigure}

    \caption{Speed-up when migrating \funcs \textnormal{(We deploy 8~MPI processes fragmented across two VMs and migrate 4 at runtime. We report the speed-up compared to not migrating.)}}
    \label{fig:eval:migration}
\end{figure}

This experiment measures the benefit of migrating \funcs at runtime. As baselines, we run a compute-bound LAMMPS simulation, and a network-bound \emph{all-to-all} kernel that performs synchronisation over a vector in a loop. To migrate a function, \sys must guarantee that there are no messages in-flight, and it uses calls to barrier synchronisation points to check for migration opportunities.

\autoref{fig:eval:migration} shows the speed-up achieved when migrating. We force the scheduler to over-fragment the jobs, and then migrate at 20\%, 40\%, 60\%, or 80\% of execution. For reference, we also include the speed-up for a co-located deployment (\textsf{1~VM}).

For a network-bound kernel, over-fragmenting has a high cost: the speed-up for \textsf{1~VM} is 7.5$\times$. By migrating after 20\%, 40\%, 60\%, and 80\%, of execution, we achieve speed-ups of 3.5$\times$, 2.7$\times$, 1.7$\times$, and 1.2$\times$, respectively. We conclude that it is always worth to migrate \funcs at runtime for network-bound applications.

For a compute-bound kernel, over-fragmenting has a lower cost: the speed-up for \textsf{1~VM} is 1.2$\times$. This is because the fragmentation splits 4~processes in 1~VM, and 4 processes into another, which means that there is substantial intra-VM messaging. By migrating after 20\%, 40\%, and 60\% of execution, we achieve speed-ups of 10\%, 8\%, and 5\%, respectively. When migrating after 80\% of execution, the costs of migrating outweigh its benefits, achieving a slow-down of 5\%. LAMMPS has large code and data sections, which leads to larger \func snapshot, increasing the cost of migration.

%% file: sections/related.tex
\section{Related Work}
\label{sec:rw}

\mypar{Scientific applications in the cloud} Nowadays, all major cloud providers offer targeted solutions to support scientific applications in the cloud~\cite{GenomicsCloud,AzureHPCWebsite,AzureGenomics}. To schedule and execute these applications, providers deploy batch scheduling solutions~\cite{AzureBatchWebsite, AWSBatchWebsite} inspired by HPC batch schedulers, and there is related work focuses on optimising scheduling decisions~\cite{ScientificScheduling1}. Instead, \sys{} focuses on utilising scheduled resources more efficiently---an orthogonal problem to better scheduling decisions. Recent work on scheduling for deep learning training on shared GPU clusters~\cite{AntMan:2020} uses traits of the scheduled resources to improve scheduling decisions, and we plan on exploring this in the future.

\mypar{Shared memory and message passing in the cloud} Hoplite~\cite{Hoplite:2021} uses well-known collective communication algorithms for building fault-tolerant task-based distributed systems. \sys adopts a similar approach: it focuses on dynamic group membership without considerations of fault-tolerance; Ray~\cite{Ray:2018} is a distributed system that unifies task-parallel and actor-based computations in a single interface. It offers transparent state and message passing irrespective of the distribution, together with transparent unlimited scaling. \sys focuses on sharing resources among multiple users more efficiently.

\mypar{Fine-grained distribution in serverless} \sys{} borrows techniques from research on serverless runtimes to allow fine-grained scheduling and distribution of scientific applications. Faasm~\cite{Faasm:ATC:2020} and Cloudburst~\cite{Cloudburst:2020} add state to serverless functions, but do not provide generic shared memory required for multi-threading, nor do they support message passing; PLASMA~\cite{PLASMA:2020} supports annotations to specify elasticity constraints but only within an actor-based programming model; Crucial~\cite{Crucial:2019} uses Java concurrency abstractions to execute serverless functions, but it lacks general multi-threading capability. The lack of communication primitives in serverless has been recognised as a limitation: SAND~\cite{SAND:2018} includes a message bus, but does not provide an associated programming model to support collective communication.

\mypar{Checkpointing and Migration} CloudScale~\cite{CloudScale:2011} automates fine-grained elastic resource scaling in a shared (multi-tenant) cluster. CloudScale also uses migration to correct scheduling or scaling issues. \sys{} could benefit from being used together with CloudScale, as it allows for finer-grained resource management, allowing CloudScale to operate at a thread/process level rather than at a VM one.  CRIU~\cite{CriuWebsite} is a software tool for checkpointing and restoring processes in userspace. \sys{}'s use of WebAssembly means that snapshot can be obtained more easily without using CRIU or other similar tools such as DMTCP~\cite{DMTCPGithub}. 





%% file: sections/conclusion.tex

\section{Conclusions}
\label{sec:concl}

Cloud computing offers on-demand parallelism that is well-suited to scientific workloads. Today's cloud services for scientific applications execute workloads on dedicated VMs, which reduces providers' abilities to re-allocate underused resources. Serverless cloud computing promises to overcome these issues through the fine-grained allocation of tasks to resources.

We have described \sys, a new cloud runtime that transparently distributes scientific workloads at a fine granularity while remaining compatible with the popular OpenMP and MPI APIs. \sys relies on \funcs, which permit the arbitrary distribution of threads and processes. Its scheduler allocates \funcs in a flexible fashion, allowing \funcs to exchange messages asynchronously and supporting a distributed shared memory implementation.




